# Probing Lithium Ion Transport at Individual Interfaces

Kartik Venkatraman[1*], Yongqiang Cheng[2], Alexandra Moy[3,4], Jordan A. Hachtel[1], Michael J. Zachman[1], Abinash Kumar[1], Olivier Delaire[5], Jeff Sakamoto[3,4], Miaofang Chi[1*]

[1]Center for Nanophase Materials Sciences, Oak Ridge National Laboratory

[2]Neutron Scattering Division, Oak Ridge National Laboratory

1 Bethel Valley Rd, Oak Ridge, Tennessee 37830, U.S.A.

[3]Department of Mechanical Engineering, University of Michigan

[4]Department of Materials Science and Engineering, University of Michigan

1221 Beal Ave, Ann Arbor, Michigan 48109, U.S.A.

[5]Department of Mechanical Engineering & Materials Science, Duke University,

144 Hudson Hall, Durham, North Carolina 27708, U.S.A.

**Ion transport across solid-solid interfaces is often slower than through the bulk of a material, impeding the charge and discharge rate of batteries[1–4]. Designing highly conductive interfaces is challenging due to the need to probe ion conduction at individual interfaces and correlate it with the local structure. In this study, we address this challenge by enabling the simultaneous measurements of local Li-dominated optical phonons and ion distributions, using high-energy and high-spatial-resolution spectroscopy in a scanning transmission electron microscope (STEM). Further, this method allows for a direct correlation of ion conduction with interfacial structures identified by STEM imaging. By examining diverse individual interfaces of $LiCoO_2$, we reveal the sensitivity of ion conduction to interface atomic-scale structure and chemistry. Our method enables correlative analysis of ion transport behavior, atomic and band structures, and can serve as a robust experimental approach for identifying interface structures that offer high conductivity and cyclability for batteries.**

The growing incorporation of electric vehicles and portable smart devices into our lives elevates the requirements for battery safety, performance, and longevity[5,6]. A key metric in gauging the performance of current and next-generation batteries is the charging/discharging rate, which is primarily associated with ionic conduction through the system – not only within the bulk of each individual component, but more critically, across the interfaces between the components. Indeed, the design and synthesis of single-phase fast ion conductors, including both electrodes and electrolytes, is now achievable[7–10]. The largest hurdle remaining for enhancing device charging rates is conduction across interfaces between electrodes and electrolytes, and across grain/phase boundaries in polycrystalline conductors and composites, where sluggish ion transport is commonly observed[1,2,11–13]. Designing interfaces with facile ion transport is therefore critical for improving the charging rates. However, conventional ion conduction measurement techniques such as electrochemical impedance spectroscopy provide measurements averaged over the microstructure of a specimen[14,15], however, making it challenging to isolate contributions from individual interfaces. Our current understanding of ion transport behavior at single interfaces relies mainly on DFT calculations, which often use symmetric GB structural models. These calculations highlight the dependence of ion transport on the interfacial lattice structure. Thus, high energy grain boundaries, which are often the majority of all grain boundaries, would behave differently from low energy symmetric ones. To design the best GB configuration for high conductivity and cyclability, the correlations between atomic-scale interfacial structure and its properties should be understood. A technique that can probe the structure, conductivity, and electronic structure with high spatial resolution is essential.

High spatial resolution methodologies such as scanning transmission electron microscopy (STEM) have been commonly deployed to directly unveil the structure, chemistry, and electronic band structures of individual interfaces that are associated with dendrite growth[16,17]. While conventional STEM methodologies are insensitive to ion transport phenomena, modern STEM monochromators now provide sufficient energy resolution to enable electron energy-loss spectroscopy (EELS) vibrational modes while maintaining Ångström-scale spatial resolution[18,19]. This technique has been demonstrated to detect localized vibrational responses of defects and nanoscale features, and the concentration of lithium ions in ion conductors [19–22]. However, utilizing vibrational EELS to study ion transport behavior has not yet been explored. While vibrational spectroscopy does not provide a direct measurement of ionic conductivity, optical phonons have long been theorized to assist in the diffusion of mobile species in solid ion conductors[23] and, as such, provide a proxy for conductivity measurements. Indeed, the energy materials community has recently exhibited a renewed focus on understanding the strong correlation between the low-lying transverse optical phonons associated with mobile ion motion and the activation energy for ion transport[24–29]. These studies have provided insights into the ion conduction mechanisms of many newly developed fast ion conductors largely by utilizing inelastic neutron scattering (INS). INS-based vibrational spectroscopy faces the same challenges as conventional ion conduction measurement techniques, however, in its inability to characterize ion transport at individual interfaces. High spatial-resolution vibrational STEM-EELS can thus directly access critical individual interface information.

Here, we demonstrate how monochromated STEM-EELS can be used to directly measure localized vibrational responses at interfaces of ion conducting materials, using grain and phase boundaries in a polycrystalline $LiCoO_2$ (LCO) as prototype systems. We develop a method for separating the contribution of the Li-dominated optical (LDO) phonons responsible for Li-ion transport from the remaining the vibrational response. By combining the structure and local Li deficiency of the boundary revealed by conventional STEM and EELS with the measured vibrational response, we measure the increase in activation energy for Li-ion transport and the corresponding reduction in ionic conductivity at the grain boundary.

**Identifying the LDO phonons**

LCO crystallizes in a layered structure with a $R\bar{3}m$ space group, wherein $CoO_6$ and $LiO_6$ octahedra form parallel layers, making it a two-dimensional ion conductor with Li ions diffusing along {001} planes. Figure 1a shows an atomic-resolution integrated center-of-mass (iCoM) STEM image of an LCO specimen oriented along [110]. Strong signals are observed for the Li, Co, and O atomic columns, as reported previously[30], and the Li ion channels along {001} planes between $CoO_6$ octahedra are clearly observed (the green arrows indicate the direction of Li ion motion during its transport). Grain boundaries were created by sintering and densifying LCO powders, forming a polycrystalline LCO pellet (See method section and Extended Data Figure 1).

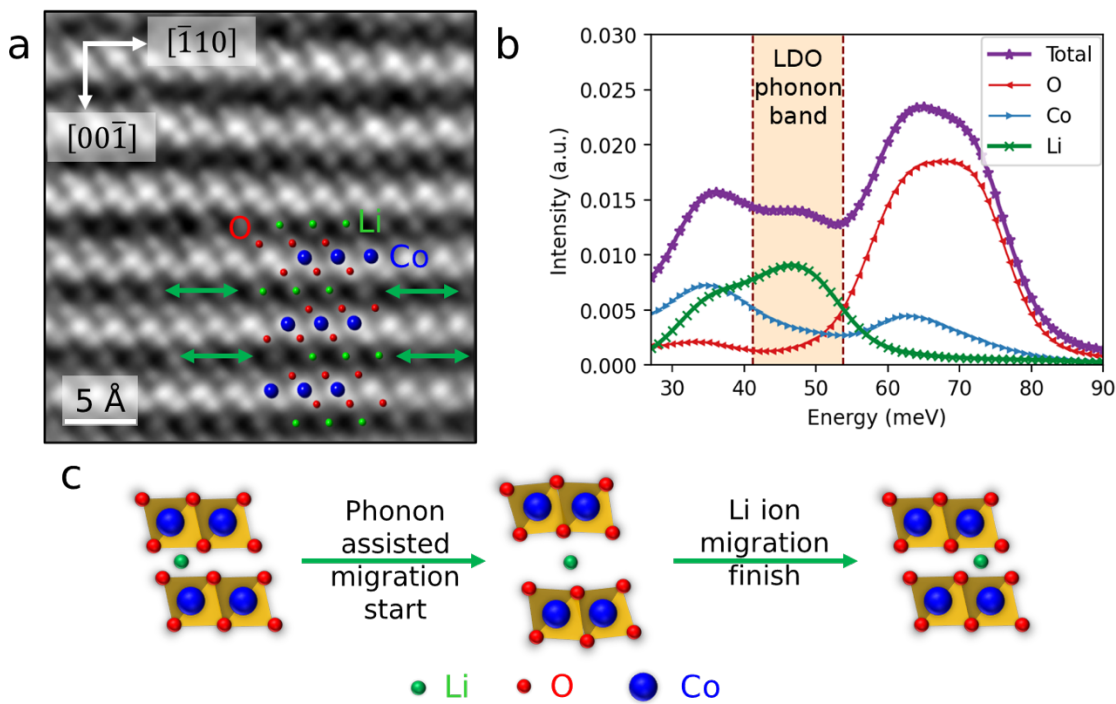

**Figure 1. Atomic structure and phonon density of states of LiCoO$_2$.** (a) Atomic resolution STEM image of [110] LiCoO$_2$ (LCO) showing Li, Co and O atomic columns (overlayed schematic shows Li (green) and Co (blue) columns as balls, and O columns as red octahedra). Green arrows indicate the direction of motion of the Li ions in the optical phonons that affect its transport. (b) Density functional theory (DFT)-calculated total phonon density of states (PDOS) for stoichiometric LCO and partial contributions from each element. The orange shaded band highlights the portion of the PDOS that primarily corresponds to Li-dominated optical (LDO) phonons. (c) A schematic representation of the LDO phonon assisted Li-ion transport in a LCO specimen oriented along [110].

To gain an intuitive understanding of the features in the vibration spectrum of LCO, which has a complex crystal structure with three atomic species, we use DFT to calculate its total phonon density of states (PDOS), along with the partial contributions of each element, as presented in Figure 1b. The region from 60 – 80 meV shows peaks in the partial PDOS of O and Co, while 40 – 55 meV shows a peak mainly in the Li partial PDOS, and < 40 meV shows peaks mostly in the partial PDOS of Li and Co. Thus the DFT calculations suggest that the best spectral range to probe the LDO phonons will be around 40 – 55 meV, consistent with previous studies[31], which will therefore be the energy region of interest for the experimental analysis of Li-ion transport by vibrational EELS.

Based on these calculations, a schematic representation of LDO phonon-assisted ion transport in an LCO crystal oriented along [110] was generated and is shown in Figure 1c. When the Li ion attempts to migrate, it experiences an energy barrier associated with the $CoO_6$ octahedra, which impedes its motion to a neighboring lattice site. The relatively large Li displacement associated with the LDO phonons then assists the Li ion in crossing this barrier, after which it proceeds to diffuse to the next equilibrium position unimpeded. The orange shaded region in Figure 1b shows the spectral range for the LDO phonons associated with this motion in the PDOS, according to the DFT results above. An animation of the ~47 meV phonon corresponding to this schematic representation can be found in Extended Data Figure 2.

**Isolating the LDO phonon contribution**

Since it is the LDO phonons that are associated with Li-ion transport in our material, it is necessary to isolate their contribution from other vibrational signals in the spectrum. To do this, we must consider electron-phonon interaction mechanisms, predominantly dipole and impact scattering[32]. Dipole scattering is associated with long-range Coulomb interactions and gives rise to spectral features that are delocalized in real space[33–35]. Impact scattering, on the other hand, is associated with short range interactions with core electrons and yields a signal that can be localized down to the atomic scale[18,36]. Normally, EELS yields both dipole and impact scattering contributions, obscuring the origin of individual peaks[37]. To separate them, we compare a traditional transmitting acquisition geometry with an 'aloof' acquisition geometry (schematically shown in Figure 2a). In the aloof mode, the probe passes close to the sample without intersecting it, enabling delocalized dipole scattering while preventing localized impact scattering. By comparing the aloof signal to the transmitting signal, we can directly extract a localized contribution to the spectrum (see Supplementary Information Section 1 for details).

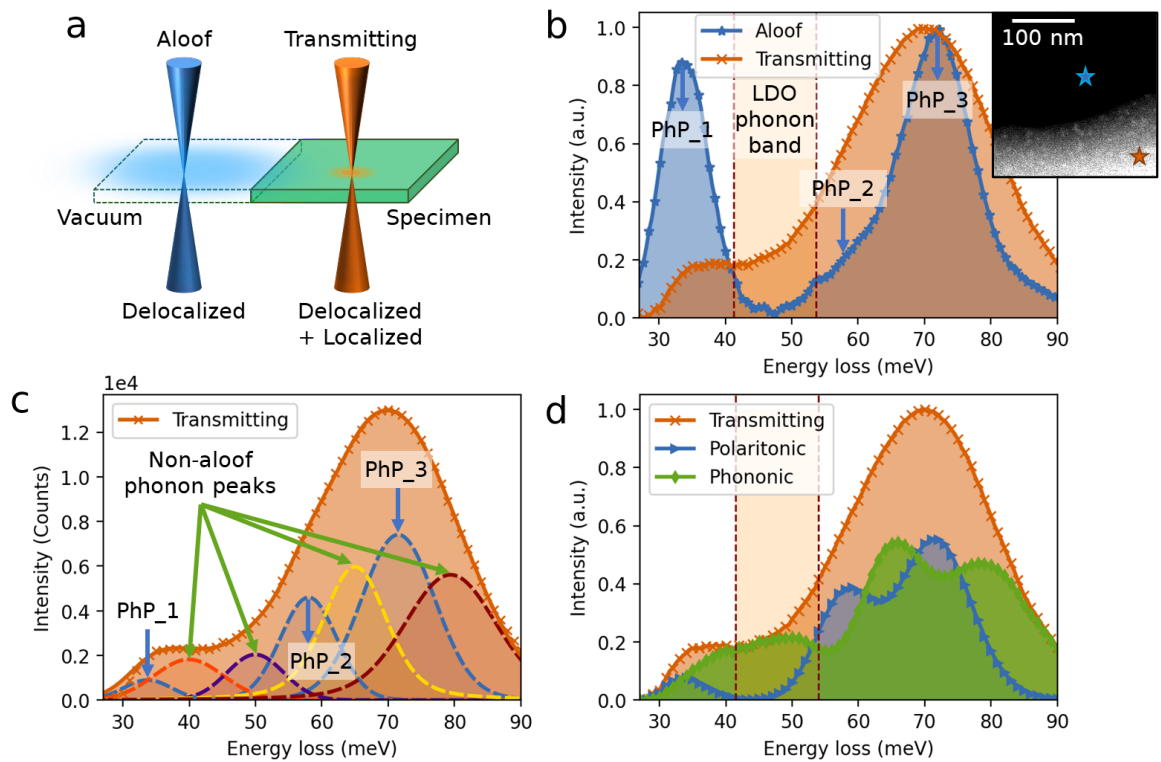

**Figure 2. "Aloof" and "Transmitting" EELS results.** (a) Schematic representation of aloof vs. transmitting EELS acquisitions. (b) Experimental aloof and transmitting vibrational spectra from [110] LCO showing differences between delocalized and localized signals. Inset: STEM image showing acquisition locations. (c) Voigt peak fitting of the transmitting vibrational spectrum. The peaks associated with aloof spectral features were added to give the polaritonic component, while the non-aloof phonon peaks were added to give the phononic component of the transmitting spectrum. (d) Polaritonic and phononic components of the transmitting vibrational spectrum. The orange shaded region highlights the spectral range associated with Li ion motion, and that it corresponds to the localized phononic signal.

Figure 2b shows aloof and transmitting EELS probe positions relative to an LCO specimen, oriented along the [110] axis, and their respective spectra. The aloof spectrum shows two sharp peaks at 34 and 72 meV and a weak shoulder between 50 and 65 meV (marked by an arrow), which are associated with other phonon polaritons (PhP_1, PhP_3, and PhP_2, respectively). See Supplementary Information Section 2 for further information on identification of aloof spectral features. Phonon polaritons are highly delocalized excitations commonly observed in vibrational EELS measurements of ionic materials[38–42]. The polaritonic signals are also present in the transmitting vibrational spectra, as shown in Figure 2b, albeit with different relative intensities and peak energies. This is due to convolution with other vibrational signals, which are responsible for changing the spectral response. The most critical change in the transmitting spectrum is the additional intensity observed between 40 and 70 meV. Recall from Figure 1 that the spectral range from 40 to 55 meV is associated with LDO phonons, which indicates that these phonons may be directly probed with highly localized impact scattering measurements and therefore can be measured at high resolution in real space, such as at individual grain boundaries or interfaces.

The contributions from phonon polaritons to the transmitting spectrum complicate spectral interpretation as they are strongly affected by specimen geometry and orientation[43]. To qualitatively separate polaritonic effects from those due to structural and chemical variations, the aloof contribution was removed from the transmitting spectrum by a Voigt decomposition. Three peaks were used to fit the aloof spectrum, while four additional peaks were required to fit the transmitting spectrum, as shown in Figure 2c (for further details see Methods and Extended Data Figure 3). The Voigt peaks associated with aloof spectral features were added to provide the "polaritonic" component of the transmitting spectrum, while the remaining peaks required for the fit were added to give the "phononic" component in Figure 2d. Based on the previous descriptions of the LCO total and partial PDOS, the phononic component features between 60 and 75 meV are primarily associated with vibrations of the $CoO_6$ octahedra while those between 40 and 55 meV are associated with LDO phonons, i.e., the phonon modes responsible for Li-ion transport. Thus, this separation of the polaritonic component from the transmitting spectrum helps to highlight the contribution from the LDO phonons, which, in turn, helps to characterize Li-ion transport across a nanoscale grain boundary.

**Change in the activation energy for Li-ion transport at a nanoscale grain boundary**

With a method established to measure the LDO phonons, we explored the nanoscale structural and vibrational properties of grain boundaries in LCO to reveal changes in local ionic conductivity. While we have analyzed various types of grain boundaries using vibrational EELS for phonon signals and core-loss EELS for Li ion concentration distribution, comparisons will be discussed later in the manuscript. Here, we focus on using a representative grain boundary from a polycrystalline LCO sample to demonstrate our method for deducing the local ion transport characteristics. This grain boundary is a high-energy GB composed of random surfaces from two randomly oriented grains, which is an atypical GB in ceramic solid electrolytes[17]. A STEM image of this representative grain boundary (enclosed by black dash-dot lines) in a polycrystalline LCO specimen is shown in Figure 3a. Like most grain boundaries in polycrystalline ion conductors, this grain boundary is not a small angle boundary or composed of two low-index surfaces[17]. It consists of a {001} surface in Grain 2 paired with a high-index surface in Grain 1. In our experiment, we aligned the [110] axis of Grain 2 parallel to the electron beam, and as a result, the [110] axis of Grain 1 is not parallel to the beam. EELS line scans were performed across the grain boundary along the green line shown in Figure 3a. Core-loss spectra from the two extremes of the line scan (Grain 1 and Grain 2) and the grain boundary (Boundary) are shown in Figure 3b. The spectra are dominated by the Li K-edge peak at 61 eV overlapping with Co $M_{2,3}$-edge (61-68eV) with additional weak signals at lower energies for the Li K-edge (54.5 – 56.5 eV) and Co $M_{2,3}$-edge (58 – 60 eV)[44]. The core-loss EELS line profile shows that the Li K-edge peak decreases in intensity while the Co $M_{2,3}$-edge increases at the grain boundary (Figure 3c). The Li:Co ratio decreases by ~27% at the boundary, as evidenced by the ratio of their edge intensities (Extended Data Figure 4), demonstrating that the chemistry changes at the grain boundary along with the structure.

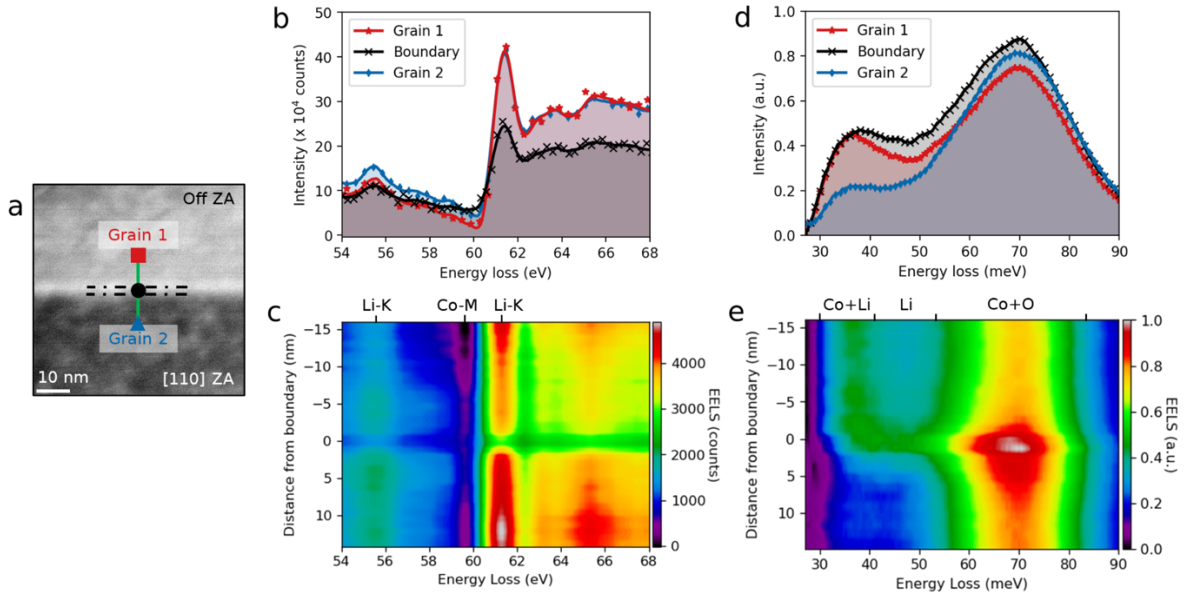

**Figure 3. Composition and vibrations across a grain boundary in polycrystalline LiCoO$_2$.** (a) STEM image of a grain boundary in bulk polycrystalline LCO where the STEM-EELS line scan was performed. [110] of Grain 2 is aligned parallel to the electron beam, while [110] of Grain 1 is not. (b) Representative core-loss EEL spectra from Grain 1, Boundary, and Grain 2 showing the Li K and the Co $M_{2,3}$ ionization edges. (c) Spatial variation profile of core-loss EELS data along the line scan showing localized reduction of Li K-edge features and enhancement of Co $M_{2,3}$-edge features. (d) Transmitting vibrational EELS from Grain 1, Boundary, and Grain 2 from the same locations as the core-loss spectra, normalized to the ZLP intensity. (e) ZLP-normalized spatial variation profile of vibrational EELS data along the line scan.

Vibrational spectra acquired at the same locations as the core-loss spectra above are shown in Figure 3d, with the vibrational EELS line profile (ZLP-normalized) shown in Figure 3e. The inelastic scattering probability in the grain boundary is higher than the grains due to the change in stoichiometry observed in core-loss, resulting in an overall increase in phonon intensity at the boundary after ZLP normalization[45]. The effect can be seen clearly in the high-energy peak between 60 and 80 meV, which consists of contributions from polaritons and localized $CoO_6$ octahedral vibrations (as shown in Figure 2d). This peak retains the same spectral shape at all locations but is enhanced in intensity at the grain boundary, indicating an increased cross-section that could be due either to a higher $CoO_6$ density or to this increase in the phonon scattering probability.

Differences in the spectra of the grains are noticeable as well. For instance, the lower energy feature between 30 and 40 meV is significantly less intense in Grain 2 than Grain 1. The ~34 meV phonon polariton majorly contributes to this feature, and such differences stem from the fact that the excitation of phonons and polaritons heavily depend on their propagation directions relative to the electron beam[43]. As the 34 meV polariton propagates along the [110] direction, its contribution is suppressed in the case of Grain 2, where the [110] axis is parallel to the electron beam, but not in Grain 1, where some component of [110] is perpendicular to the beam.

The most important component of these spectra for our purposes, however, is the "valley" between 40 and 55 meV, where the LDO phonons are present. In addition to the subtle differences between Grain 2 and Grain 1, which are likely due to orientation effects, the normalized valley intensity is enhanced at the grain boundary compared to the grains in the spectral region of the LDO phonons (40 – 55 meV), demonstrating that even though the relative Li concentration is reduced in the grain boundary, the Li vibrations still play a dominant role.

The detailed characters of the grain boundary vibrational responses can be understood by decomposing the spectra into their polaritonic and phononic components using the Voigt decomposition, as shown in Figures 4a and 4b. The out-of-plane polariton (34 meV peak in Figure 4a) is dominated by orientational effects, and the grain boundary peak appears to be a midpoint between the two grains as expected. Conversely, the higher energy polariton peaks (between 58 and 72 meV) are nearly equivalent in the grains but differ at the grain boundary, indicating a fundamentally different vibrational response at the boundary. As for phononic components (Figure 4b), the grain boundary signal is enhanced overall relative to the grains, indicating that impact scattering is responsible for the enhanced phonon cross-section in Figures 3d and 3e. The phononic signal at ~40 meV varies significantly with grain orientation so that it is likely associated with an out-of-plane phonon.

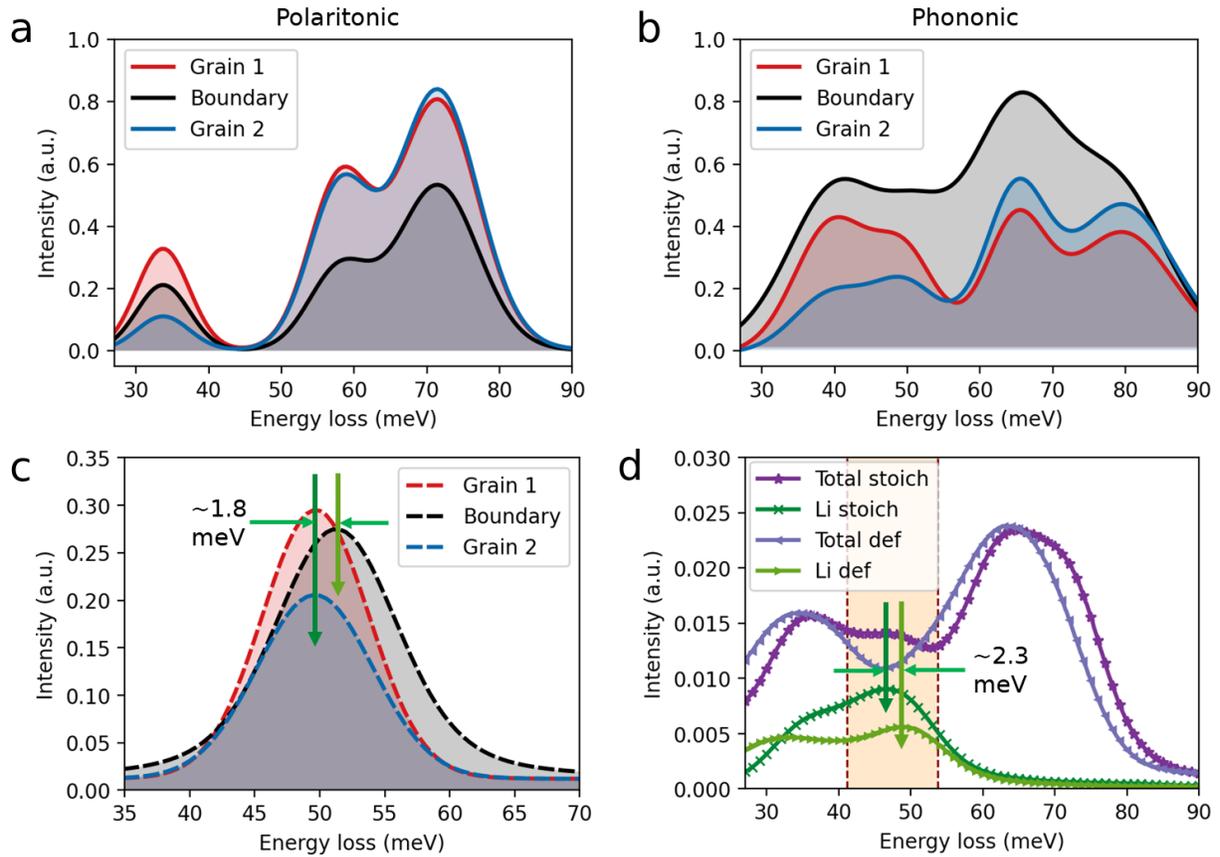

**Figure 4. Polaritonic and phononic contributions to vibrational EELS across an individual nanoscale grain boundary.** (a, b) Representative polaritonic (a) and phononic (b) components from Voigt decomposition of transmitting vibrational spectra in Figure 3d. (c) The Voigt peak associated with the LDO phonons at ~50 meV from Grain 1, Boundary and Grain 2 showing a ~1.8 meV blue shift in peak energy position at the grain boundary. (d) Comparing the total PDOS for stoichiometric LCO (purple) with Li-deficient LCO (violet). Partial contribution from Li for pristine LCO (dark green) and deficient LCO (lime green) shows similar shift to EELS.

More critical is the peak at ~50 meV, which is most associated with the LDO phonons. While the intensity of the phonoic component can not be easily interpreted due to the mixed contribution from projected density of states and the dynamic form factor (see Supplementary Information), shifts in the spectral position of the phononic Voigt peaks can be affected only by the density of states, and are representative of the local structural and chemical variation in the specimen. A clear blue shift of ~1.8 meV relative to that of the grains can be seen in grain boundary peak (Figure 4c). Since the LDO phonon frequency is directly proportional to the activation energy for ion transport, a blue shift thus implies an increased activation energy at the grain boundary.

To better understand the origin of this blue shift, we consider the contribution of the measured Li deficiency by calculating the PDOS with one-third of the Li atoms removed from stoichiometric LCO, roughly matching the change in chemistry at the grain boundary (~27% less Li relative to Co). A comparison of the total PDOS for stoichiometric LCO and Li-deficient LCO, along with the partial contribution from Li, is shown in Figure 4d. We observe that the Li peak at ~47 meV blue shifts by ~2.3 meV for the Li-deficient structure, meaning Li-deficiency induced local structure and chemical changes are likely the primary cause for the experimentally measured blue shift of the ~50-meV transmitting phonon peak at the grain boundary. It's noted that ion vibrations are closely linked to the coordination of the ions within the lattice, at high-angle grain boundaries such as this where the local lattice structure substantially deviates from the bulk structure, the divergence between experimental and calculated results using a simplified structure model (~0.5meV) should be attributed to the lattice structural deviations. The smaller blueshift in experiment value compared to the calculated one suggest that this grain boundary structure may promote rather than impede ion conduction. These results show that EELS-vibrational signals take account of the contributions from both local chemistry and structure, and our method can directly measure phonon modes related to ion conductivity across interfaces at the nanoscale, regardless of their complex atomic lattice structures.

An increase in energy of approximately 1.8 meV for LDO phonons at around 50 meV would result in a qualitative increase of approximately 7% in the activation energy for Li-ion transport at the grain boundary, based on Wakamura's empirical equition[23]. This increased activation energy indicates a decrease in the mobility of lithium ions at this specific grain boundary compared to the grains. Now we have obtained local diffusivity, which can now be use to estimate the ion conductivity by combining the mobile ion concentration that were quantified by core-loss EELS. As we quantify ~27% less Li relative to Co at this grain boundary from core-loss EELS, a total conductivity drop of ~29% can be estimated based on the combination of Einstein and Arrhenius equations[46] (further details shown in Supplementary Information Section 3). These findings challenge the conventional understanding that grain boundaries typically enhance ion diffusion, especially in relation to ideal symmetry boundaries. It seems that this understanding does not apply to high-angle, high-energy grain boundaries.

The method presented in this study is applicable to interfaces of any crystal structure and phases. Extended Data Figures 9-11 demonstrate the correlative vibrational and core-loss EELS analysis of various grain boundaries, including a high-energy and high-angle grain boundary, a highly twisted grain boundary with two grains in close proximity to a high-index zone, and a phase boundary with significantly different Li concentrations between adjacent grains. Detailed discussions of these grain boundaries are provided in their respective captions. The results reveal substantial variations in Li vibrational signals and Li concentration at the boundaries, underscoring the importance of comprehensively understanding microscopic parameters such as atomic structure, chemistry, ion transport, and band structure in future efforts to identify high-performance interfaces for batteries.

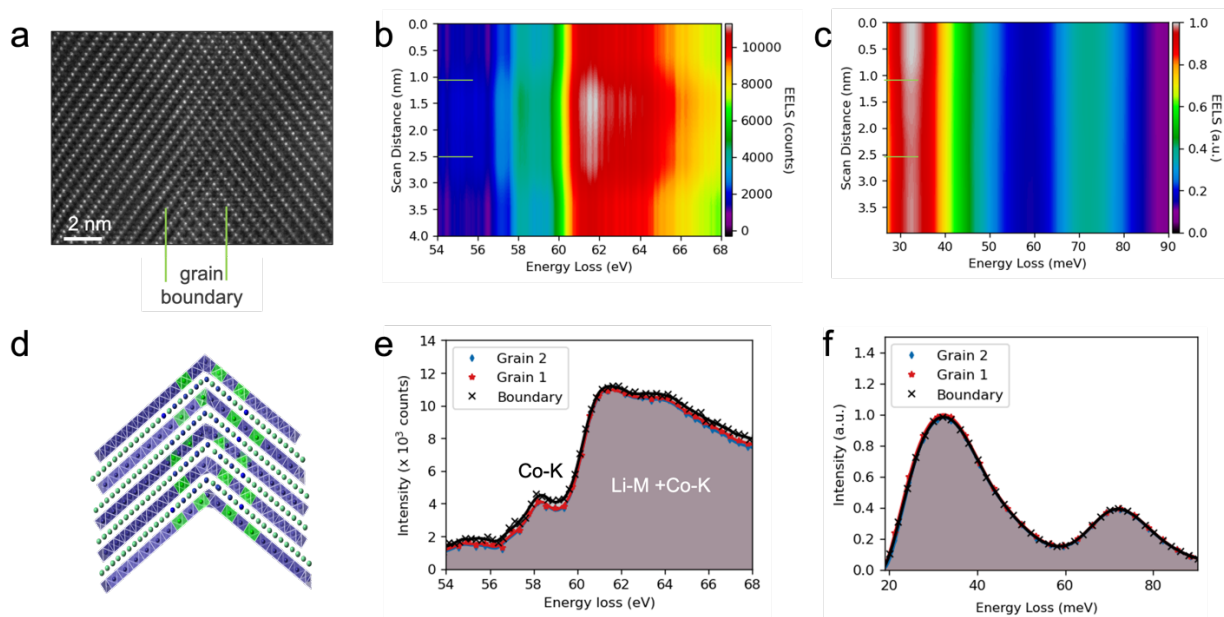

**Figure 5. Correlated vibrational and core loss EELS of a symmetric grain boundary.** HAADF-STEM image (a), two-dimensional maps of core-loss (b) and low-loss (c) EELS, simplified schematic of its GB structure showing cation mixing (d); representative core-loss (e) and low-loss (f) integrated spectra of the two grains and the grain boundary. (a, b) green lines in (a-c) marks grain boundary region.

In an attempt to compare our experimental results with existing theoretical calculations, we utilized Pulsed Laser Deposition (PLD) to grow a symmetric grain boundary, as depicted in Figure 5. The grain boundary exhibits a highly twisted near-Σ2 structure, as evident from the reduced angle in the projection view [47,48]. The grain boundary is not a sharp in chemistry and structure, as evidenced in the core loss map which is ~1.6nm, corresponding to 3-4 u.c. thickness based on signal variation in Figure 5b. Core-loss analysis reveals a minimal increase in the Co/Li ratio at the grain boundary core, as evidenced by the slightly larger signal at ~58.2 eV (Figure 5d). The vibrational signal at the grain boundary is identical to that of the grain (Figure 5c and 5e), suggesting that the activation energy of Li ions remains unchanged despite the higher Li concentration. This finding contradicts previous theoretical calculations suggesting enhanced ion conduction along and across Σ2 grain boundaries[47]. By closely examining the atomic structure of this grain boundary, we identified the potential cause of this discrepancy. As shown in the figure, there is significant cation mixing of Li/Co observed at the grain boundary core. This cation mixing could strongly hinder the Li percolation pathway, which explains why this grain boundary does not exhibit a lower Li activation energy as predicted regardless of a slightly increased Li concentration. These findings highlight the advantage of utilizing correlative analysis of vibrational EELS and imaging in a STEM.

**Conclusions**

By combining core-loss and vibrational EELS in a monochromated STEM, we directly measure and quantify variations in ion concentration and vibrational signals at interfaces. This enables the assessment of ion conduction behavior at individual interfaces. Conducted within a STEM, this approach allows for the correlation of ion concentration, atomic and band structure, and local phonon modes, which collectively define interfacial ion transport and dendrite growth resilience—critical factors for battery applications. Through the analysis of diverse interfaces, we gain insights into the sensitivity of ion conduction to interface atomic scale structural and chemical configurations. Future systematic studies utilizing this method will facilitate the identification of interfaces with optimal performance. Importantly, this versatile method extends beyond battery materials and can be applied to ion conducting interfaces in diverse energy applications, such as solid-state fuel cells, ion exchange membranes, and sensors, among others.

# Methods

**Synthesis of LiCoO$_2$ pellet:**

LiCoO$_2$ powders (EQ-Lib-LCO, MTI, California, USA) were cold-pressed into billets at ~175 MPa at room temperature in a half-inch diameter stainless steel die. The cold-pressed billet was then sintered and densified using a rapid induction hot-press at 60 MPa for 60 minutes at 900°C under a flowing argon atmosphere. The hot-pressed billet was then cut into pellets of ~1.7 mm thickness using a diamond saw then manually ground using 1200 grit and 2500 grit sandpaper.

**TEM specimen preparation:**

A 2 x 1 x 1 mm piece was cut from the bulk polycrystalline LiCoO$_2$ pellet using a SYJ-150 low speed diamond cutting saw by MTI Corporation, after which it was mounted on a stub using specimen mounting wax from Agar Scientific. It was then mechanically polished to ~10 μm thickness using the MultiPrep Polishing System from Allied High Tech Products, Inc. An alcohol based lubricant was used during the mechanical polishing step. The polished sample was transferred to a 3 mm Mo slot grid using M-Bond 610 adhesive, followed by ion mill preparation to thin the sample to typical TEM specimen thickness using the Fischione Model 1010 TEM Mill system. Further surface cleaning was performed using the Fischione Model 1040 NanoMill TEM specimen preparation system. All ion milling was performed with the specimen at cryogenic temperatures. The specimen was baked at 140°C in vacuum for 12 hours prior to introduction into the microscope to remove volatile hydrocarbons and reduce contamination.

**STEM Imaging:**

The atomic-resolution image in Figure 1 is acquired by integrated Center of Mass (iCoM) STEM imaging, which allows light and heavy atoms to be imaged simultaneously[49,50]. 4D-STEM data was acquired on a Nion UltraSTEM 100 at Oak Ridge National Laboratory (ORNL), equipped with a Hamamatsu ORCA CMOS detector and operated at 100 kV with a convergence semi angle of ~32 mrad. The central 128 × 128 pixels of the detector were used to increase the data acquisition rate to $10^{-3}$ s per probe position and reduce file sizes. 128 × 128 real-space probe positions were used, for total acquisition times of approximately 1 min or less. The center of mass (CoM) of each acquired diffraction pattern was calculated to provide a two-dimensional CoM vector map. The mean CoM signal was subsequently subtracted from this map. An integration step was finally performed to produce iCoM images.

STEM images in Figure 2 and Figure 3 are high-angle annular dark-field (HAADF) images acquired on the ORNL Nion High Energy-Resolution Monochromated EELS STEM (HERMES) 100 operated at 60 kV with a convergence semi angle of 35 mrad.

**EELS measurements:**

STEM EELS analysis on the specimen was performed using the ORNL Nion HERMES 100 operated at 60 kV with a convergence semi angle of 35 mrad and a collection semi angle of 25 mrad. EELS information was acquired as line scans across specific grain boundaries, wherein energy-loss spectra were recorded at every pixel along a line, or as 'point-and-shoot' spectra from different regions in and around the specimen. One of the grains forming the boundary was oriented along the [110] axis before all EELS acquisitions. Energy-dispersions of 0.4 meV/ch., 0.05 eV/ch., and 0.1 eV/ch. was used to record vibrational-loss, valence-loss, and core-loss spectra, respectively. All spectra were calibrated by setting the maximum of the unsaturated ZLP to 0 meV, and hence, the error in measuring energy-loss was the channel width. Data processing was performed with a combination of the Gatan Microscopy Suite, Nion Swift and custom Python codes. Relative thickness (thickness per inelastic mean free path-length) along the vibrational-loss and core-loss EELS linescans was estimated using the valence-loss EELS linescans in the same location.

The aloof spectrum in Figure 2b was acquired with the probe positioned ~65 nm away from the specimen surface. For all vibrational spectra, parametric two-window background subtraction using the power-law model was performed to isolate the energy-loss features of interest from the background of the tail of the zero-loss peak (ZLP) and uncharacteristic energy-losses preceding the feature of interest. In Figure 3b, the core-loss spectra were obtained by spatially averaging over 2 nm. The solid symbols are experimental data, while the solid lines are processed data obtained using a Savitzky-Golay smoothing filter (49 points, polynomial order 5)[51]. For core-loss spectra, single window background subtraction using the power law model was performed, and the window was positioned between 43 and 45 eV, as this was the closest position to the Li K-edge onset at ~55 eV that did not yield a negative intensity in the resultant signal across the entire line scan.

**Voigt peak fitting to LCO vibrational spectra:**

A custom python script (using LMFIT package) was utilized to define Voigt functions and perform curve fitting to background subtracted vibrational spectra from LCO using non-linear least-squares minimization. Standard reduced chi-squared statistics was applied to determine the goodness of the fit, to reduce the possibility of under and over fitting. A Voigt peak has four parameters – amplitude, center, sigma, and gamma, where sigma and gamma majorly affect the width and the tail of the peak, respectively. The amplitude for all Voigt peaks was allowed to vary between 0 and infinity (inf). The initial guesses for peak amplitude and center were chosen by surveying the spectral intensity and position for every feature. The lower limit on the full-width at half-maximum (FWHM) for the peaks, which is a function of sigma and gamma, was chosen to be 8 meV based on our experimental energy-resolution, while the upper limit was chosen to be 20 meV as this was the corresponding FWHM of the two main features observed in transmitting vibrational spectra. We first performed the fitting of the aloof spectrum (Extended Data Figure 3a) and tried fitting between 2 and 4 Voigt peaks to it, obtaining reduced chi-squared values for every iteration. The final number of peaks was chosen as 3 based on the best reduced chi-squared value. The initial guesses for [amplitude, center, sigma, gamma] for the three peaks were [1.5e5, 34, 4, 1], [2.2e4, 54, 5, 1], [1.8e5, 74, 5.5, 1], while their lower and upper bounds were [0, 32, 3.1, 0.1], [0, 50, 3.1, 0.1], [0, 68, 3.1, 0.1] and [inf, 35, 6, 2], [inf, 58, 6, 2], [inf, 78, 6, 2], respectively. Optimized values for amplitude, center, sigma and gamma for each of the three Voigt peaks were obtained after the curve fitting. The optimized values for center, sigma and gamma for the three peaks were fixed, whereas their amplitude was allowed to vary, while performing the fitting of the transmitting spectra (Extended Data Figures 3b-d). Once again, we tried fitting between 3 and 5 additional peaks and chose the final number of additional peaks to be 4 based on the best reduced chi-squared value. The initial guesses for [amplitude, center, sigma, gamma] for the four additional peaks were [6.1e4, 41, 5.5, 0.75], [3.2e4, 49, 6, 0.75], [1.45e5, 65, 4.5, 0.75], [6.9e4, 81, 7, 0.75] while their lower and upper bounds were [0, 39, 3.1, 0.1], [0, 49, 3.1, 0.1], [0, 65, 3.1, 0.1], [0, 79, 3.1, 0.1] and [inf, 42, 6, 2], [inf, 52, 6, 2], [inf, 68, 6, 2], [inf, 82, 6, 2], respectively. Once more, optimized values for amplitude, center, sigma and gamma for each of the four additional Voigt peaks were obtained after the curve fitting. This process was repeated for all vibrational spectra along the line scan. For every transmitting spectrum, the Voigt peaks with the same centers as those used to fit the aloof spectrum were added to give the polaritonic component while the remaining Voigt peaks were added to give the phononic component.

**Zero-loss peak normalization of LCO vibrational spectra:**

The background of the zero-loss peak (ZLP) varies considerably along the line scan due to thickness and elastic scattering differences (Extended Data Figure 5). Features in the vibrational spectrum sit on top of this background, and for a fair comparison of spectral features we need to take the ZLP background variation into account. To do that, all background subtracted vibrational spectra were normalized to the integrated ZLP intensity[52]. The intensity between -2.5 and 2.5 meV in every raw spectrum was integrated and every background subtracted vibrational spectrum was then divided by the corresponding integrated ZLP intensity at that position to obtain the ZLP-normalized vibrational spectrum. All ZLP-normalized spectra were then scaled between 0 and 1 to obtain the presented data. Vibrational spectra and the spatial variation profile along the line scan without ZLP normalization are shown in Extended Data Figure 6.

**Density Functional Theory (DFT) calculations:**

DFT calculations were performed using the Vienna Ab initio Simulation Package (VASP)[53]. The calculation used Projector Augmented Wave (PAW) method[54,55] to describe the effects of core electrons, and Perdew-Burke-Ernzerhof (PBE)[56] implementation of the Generalized Gradient Approximation (GGA) for the exchange-correlation functional. A Hubbard U term of 3.32 eV[57] was applied to account for the localized 3d orbitals of Co. Energy cutoff was 600 eV for the plane-wave basis of the valence electrons. The total energy tolerance for electronic energy minimization was $10^{-6}$ eV. For the pristine LCO, a 4×4×1 supercell (48 Li, 48 Co, 96 O) of the unit cell was used as the starting structure. For the defective LCO, about 35% of Li vacancies (18 out of 48) were created randomly in the Li layers. Molecular dynamics (MD) simulations were then performed at 300 K for 10 ps, with a time step of 0.5 fs. The total and partial PDOS were obtained on the MD trajectories using OCLIMAX[58].

# References

1. Takada, K., Ohno, T., Ohta, N., Ohnishi, T. & Tanaka, Y. Positive and Negative Aspects of Interfaces in Solid-State Batteries. *ACS Energy Lett* **3**, 98–103 (2018).
2. Yu, X. & Manthiram, A. Electrode-Electrolyte Interfaces in Lithium-Sulfur Batteries with Liquid or Inorganic Solid Electrolytes. *Acc Chem Res* **50**, 2653–2660 (2017).
3. Xiao, Y. *et al.* Understanding interface stability in solid-state batteries. *Nature Reviews Materials 2019 5:2* **5**, 105–126 (2019).
4. Tan, D. H. S., Banerjee, A., Chen, Z. & Meng, Y. S. From nanoscale interface characterization to sustainable energy storage using all-solid-state batteries. *Nature Nanotechnology 2020 15:3* **15**, 170–180 (2020).
5. Zhao, Q., Stalin, S., Zhao, C.-Z. & Archer, L. A. Designing solid-state electrolytes for safe, energy-dense batteries. *Nat Rev Mater* **5**, 229–252 (2020).
6. Manthiram, A., Yu, X. & Wang, S. Lithium battery chemistries enabled by solid-state electrolytes. *Nature Reviews Materials* vol. 2 Preprint at https://doi.org/10.1038/natrevmats.2016.103 (2017).
7. Kamaya, N. *et al.* A lithium superionic conductor. *Nature Materials 2011 10:9* **10**, 682–686 (2011).
8. Zhang, W. *et al.* Kinetic pathways of ionic transport in fast-charging lithium titanate. *Science (1979)* **367**, 1030–1034 (2020).
9. Zhou, L. *et al.* High areal capacity, long cycle life 4 V ceramic all-solid-state Li-ion batteries enabled by chloride solid electrolytes. *Nature Energy 2022 7:1* **7**, 83–93 (2022).
10. Zhang, Z. & Nazar, L. F. Exploiting the paddle-wheel mechanism for the design of fast ion conductors. *Nature Reviews Materials 2022 7:5* **7**, 389–405 (2022).
11. Ma, C. *et al.* Interfacial Stability of Li Metal-Solid Electrolyte Elucidated via in Situ Electron Microscopy. *Nano Lett* **16**, 7030–7036 (2016).
12. Liu, X. *et al.* Local electronic structure variation resulting in Li 'filament' formation within solid electrolytes. *Nat Mater* **20**, 1485–1490 (2021).
13. Zhu, Y. *et al.* Lithium-film ceramics for solid-state lithionic devices. *Nature Reviews Materials 2020 6:4* **6**, 313–331 (2020).
14. He, X. *et al.* In situ atomic-scale engineering of the chemistry and structure of the grain boundaries region of Li3xLa2/3-xTiO3. *Scr Mater* **185**, 134–139 (2020).
15. Gaddam, R. R., Katzenmeier, L., Lamprecht, X. & Bandarenka, A. S. Review on physical impedance models in modern battery research. *Physical Chemistry Chemical Physics* **23**, 12926–12944 (2021).
16. Ma, C. *et al.* Atomic-scale origin of the large grain-boundary resistance in perovskite Li-ion-conducting solid electrolytes. *Energy Environ Sci* **7**, 1638–1642 (2014).
17. Liu, X. *et al.* Local electronic structure variation resulting in Li 'filament' formation within solid electrolytes. *Nat Mater* **20**, 1485–1490 (2021).
18. Venkatraman, K., Levin, B. D. A., March, K., Rez, P. & Crozier, P. A. Vibrational spectroscopy at atomic resolution with electron impact scattering. *Nat Phys* **15**, 1237–1241 (2019).
19. Hage, F. S., Radtke, G., Kepaptsoglou, D. M., Lazzeri, M. & Ramasse, Q. M. Single-atom vibrational spectroscopy in the scanning transmission electron microscope. *Science* **367**, 1124–1127 (2020).
20. Yan, X. *et al.* Single-defect phonons imaged by electron microscopy. *Nature 2020 589:7840* **589**, 65–69 (2021).

21. Hoglund, E. R. *et al.* Direct Visualization of Localized Vibrations At Complex Grain Boundaries. *Advanced Materials* 2208920 (2023) doi:10.1002/ADMA.202208920.
22. Lee, T. *et al.* Atomic-scale origin of the low grain-boundary resistance in perovskite solid electrolyte Li0.375Sr0.4375Ta0.75Zr0.25O3. *Nature Communications 2023 14:1* **14**, 1–14 (2023).
23. Wakamura, K. Roles of phonon amplitude and low-energy optical phonons on superionic conduction. *Phys Rev B Condens Matter Mater Phys* **56**, 11593–11599 (1997).
24. Bachman, J. C. *et al.* Inorganic Solid-State Electrolytes for Lithium Batteries: Mechanisms and Properties Governing Ion Conduction. *Chem Rev* **116**, 140–162 (2016).
25. Muy, S. *et al.* Tuning mobility and stability of lithium ion conductors based on lattice dynamics. *Energy Environ Sci* **11**, 850–859 (2018).
26. Liu, X. *et al.* Elucidating the mobility of H + and Li + ions in (Li 6.25Àx H x Al 0.25 )La 3 Zr 2 O 12 via correlative neutron and electron spectroscopy † ‡ Broader context. *Energy Environ. Sci* **12**, 945 (2019).
27. Muy, S. *et al.* Phonon-Ion Interactions: Designing Ion Mobility Based on Lattice Dynamics. (2020) doi:10.1002/aenm.202002787.
28. Gordiz, K., Muy, S., Zeier, W. G., Shao-Horn, Y. & Henry, A. Enhancement of ion diffusion by targeted phonon excitation. *Cell Rep Phys Sci* **2**, 100431 (2021).
29. Gupta, M. K. *et al.* Strongly Anharmonic Phonons and Their Role in Superionic Diffusion and Ultralow Thermal Conductivity of Cu7PSe6. *Adv Energy Mater* **12**, 2200596 (2022).
30. Zachman, M. J., Yang, Z., Du, Y. & Chi, M. Robust Atomic-Resolution Imaging of Lithium in Battery Materials by Center-of-Mass Scanning Transmission Electron Microscopy. *ACS Nano* **16**, 1358–1367 (2022).
31. Julien, JM. Vibrational spectroscopy of electrode materials for Li-ion batteries. in *Materials Science for Energy Storage* (ed. C.M. Julien, A. M. K. Z. A. Vijh.) 201–225 (Anna University Publ., Chennai, India, 2010).
32. Ibach, H. & Mills, D. L. *Electron energy loss spectroscopy and surface vibrations*. (Academic press, 2013).
33. Rez, P. *et al.* Damage-free vibrational spectroscopy of biological materials in the electron microscope. *Nat Commun* **7**, 10945 (2016).
34. Hachtel, J. A. *et al.* Identification of site-specific isotopic labels by vibrational spectroscopy in the electron microscope. *Science (1979)* **363**, 525–528 (2019).
35. March, K. *et al.* Protein secondary structure signatures from energy loss spectra recorded in the electron microscope. *J Microsc* **282**, (2021).
36. Hage, F. S., Kepaptsoglou, D. M., Ramasse, Q. M. & Allen, L. J. Phonon Spectroscopy at Atomic Resolution. *Phys Rev Lett* **122**, 016103 (2019).
37. Venkatraman, K. & Crozier, P. A. Role of Convergence and Collection Angles in the Excitation of Long- and Short-Wavelength Phonons with Vibrational Electron Energy-Loss Spectroscopy. *Microscopy and Microanalysis* **27**, 1069–1077 (2021).
38. Lagos, M. J., Trügler, A., Hohenester, U. & Batson, P. E. Mapping vibrational surface and bulk modes in a single nanocube. *Nature* **543**, 529–532 (2017).
39. Govyadinov, A. A. *et al.* Probing low-energy hyperbolic polaritons in van der Waals crystals with an electron microscope. *Nat Commun* **8**, 95 (2017).
40. Li, N. *et al.* Direct observation of highly confined phonon polaritons in suspended monolayer hexagonal boron nitride. *Nature Materials 2020 20:1* **20**, 43–48 (2020).

41. Li, X. *et al.* Three-dimensional vectorial imaging of surface phonon polaritons. *Science (1979)* **371**, 1364–1367 (2021).
42. Konečná, A. *et al.* Revealing Nanoscale Confinement Effects on Hyperbolic Phonon Polaritons with an Electron Beam. *Small* **17**, 2103404 (2021).
43. Radtke, G. *et al.* Polarization Selectivity in Vibrational Electron-Energy-Loss Spectroscopy. *Phys Rev Lett* **123**, 256001 (2019).
44. Kikkawa, J., Mizoguchi, T., Arai, M., Nagai, T. & Kimoto, K. Identifying lithium K edge anisotropy in LiCo O2. *Phys Rev B* **98**, 075103 (2018).
45. Egerton, R. F. *Electron Energy-Loss Spectroscopy in the Electron Microscope*. (Springer US, 2011).
46. Park, M., Zhang, X., Chung, M., Less, G. B. & Sastry, A. M. A review of conduction phenomena in Li-ion batteries. *J Power Sources* **195**, 7904–7929 (2010).
47. He, X., Sun, H., Ding, X. & Zhao, K. Grain Boundaries and Their Impact on Li Kinetics in Layered-Oxide Cathodes for Li-Ion Batteries. *Journal of Physical Chemistry C* **125**, 10284–10294 (2021).
48. Hiroki Moriwake, C. *et al.* First-Principles Calculations of Lithium-Ion Migration at a Coherent Grain Boundary in a Cathode Material, LiCoO2. *Advanced Materials* **25**, 618–622 (2013).
49. Lazić, I., Bosch, E. G. T. & Lazar, S. Phase contrast STEM for thin samples: Integrated differential phase contrast. *Ultramicroscopy* **160**, 265–280 (2016).
50. Lazić, I. & Bosch, E. G. T. Analytical Review of Direct Stem Imaging Techniques for Thin Samples. *Advances in Imaging and Electron Physics* **199**, 75–184 (2017).
51. Savitzky, A. & Golay, M. J. E. Smoothing and Differentiation of Data by Simplified Least Squares Procedures. *Anal Chem* **36**, 1627–1639 (1964).
52. Venkatraman, K., Rez, P., March, K. & Crozier, P. A. The influence of surfaces and interfaces on high spatial resolution vibrational EELS from $SiO_2$. *Microscopy* **67**, i14–i23 (2018).
53. Kresse, G. & Furthmüller, J. Efficient iterative schemes for ab initio total-energy calculations using a plane-wave basis set. *Phys Rev B Condens Matter Mater Phys* **54**, 11169–11186 (1996).
54. Blöchl, P. E. Projector augmented-wave method. *Phys Rev B* **50**, 17953–17979 (1994).
55. Kresse, G. & Joubert, D. From ultrasoft pseudopotentials to the projector augmented-wave method. *Phys Rev B* **59**, 1758 (1999).
56. Perdew, J. P., Burke, K. & Ernzerhof, M. Generalized gradient approximation made simple. *Phys Rev Lett* **77**, 3865–3868 (1996).
57. Wang, L., Maxisch, T. & Ceder, G. Oxidation energies of transition metal oxides within the GGA+U framework. *Phys Rev B Condens Matter Mater Phys* **73**, (2006).
58. Cheng, Y. Q., Daemen, L. L., Kolesnikov, A. I. & Ramirez-Cuesta, A. J. Simulation of inelastic neutron scattering spectra using OCLIMAX. *ACS Publications* **15**, 1974–1982 (2019).

# End notes

## Acknowledgements

This research was supported by the U.S. DOE Office of Science Early Career project FWP# ERKCZ55. Microscopy experiments were performed at the Center for Nanophase Materials Sciences (CNMS), which is a U.S. DOE Office of Science User Facility. This work was part of the US-German joint collaboration on "Interfaces and Interphases in Rechargeable Li-Metal Based Batteries" supported by the US Department of Energy (DOE) and the German Federal Ministry of Education and Research (BMBF). Financial support from the DOE under the grant number DE-ACO5-000R22275 is acknowledged.

## Author contributions

M.C. conceived the initial project. K.V. performed STEM-EELS experiments and data analysis, drafted the manuscript, and developed the EELS analysis method to isolate responsible phonon signals with help from A.K. and J.A.H. A.M. and J.S. synthesized the LCO pellet. M.J.Z. acquired and analyzed iCoM-STEM data. Y.C. performed DFT calculations. K.V., Y.C., M.J.Z., O.D., J.A.H. and M.C. were involved in discussions on the interpretation of the results. All authors were active in editing the manuscript.

## Competing interests

The authors declare no competing interests.

## Additional information

**Supplementary Information** is available for this paper.

**Correspondence and requests for materials** should be addressed to corresponding authors